\documentclass[journal,twoside]{IEEEtran}
\pdfoutput=1
\usepackage{graphicx}
\usepackage[cmex10]{amsmath}
\usepackage{amssymb}
\usepackage{fixltx2e}
\usepackage{url}
\usepackage{bm}
\usepackage{cite}
\usepackage[hidelinks]{hyperref}
\usepackage{eso-pic}
\usepackage{xcolor}

\graphicspath{{./gfx/}}

\newcommand{\D}{\displaystyle}
\newcommand{\Person}[1]{#1}
\newcommand{\U}[1]{\mathrm{#1}}
\newcommand{\Vector}[1]{\bm{#1}}  
\newcommand{\Matrix}[1]{\bm{#1}}  
\newcommand{\Transpose}{\mathrm{T}}  
\newcommand{\refEq}[1]{(\ref{#1})}               
\newcommand{\refFig}[1]{Fig.~\ref{#1}}           
\newcommand{\refSec}[1]{Sect.~\ref{#1}}          

\DeclareMathOperator{\limit}{sat}
\newcommand{\mylimit}[3]{\limit_{#2}^{#3}\left({#1}\right)}  

\hypersetup{%
    pdfauthor={Gernot Herbst},
    pdftitle={Practical Active Disturbance Rejection Control: Bumpless Transfer, Rate Limitation and Incremental Algorithm},
    pdfcreator={},pdfproducer={}
}

\begin{document}

\AddToShipoutPictureBG{%
  \AtTextUpperLeft{%
    \setlength\unitlength{1cm}%
    \put(0,1.3){\begin{minipage}[c]{18.14cm}
    \footnotesize\centering\textcolor{black!50}{%
    This is the author's version of an article that has been published in this journal. Changes were made to this version by the publisher prior to publication.\\
    The final version of record is available at} \ \ \textcolor{blue!60}{\url{http://dx.doi.org/10.1109/TIE.2015.2499168}
    }
    \end{minipage}}%
  }
  \AtTextLowerLeft{%
    \setlength\unitlength{1cm}%
    \put(-1,-1.3){\begin{minipage}[c]{20.14cm}
    \footnotesize\centering\textcolor{black!50}{%
    Copyright (c) 2015 IEEE. Personal use is permitted. For any other purposes, permission must be obtained from the IEEE by emailing \url{pubs-permissions@ieee.org}.}
    \end{minipage}}%
  }
}

\title{Practical Active Disturbance Rejection Control: Bumpless Transfer, Rate Limitation and Incremental Algorithm}

\author{Gernot Herbst\ \href{https://orcid.org/0000-0002-4638-5378}%
    {\raisebox{-0.3pt}{\includegraphics[height=9pt]{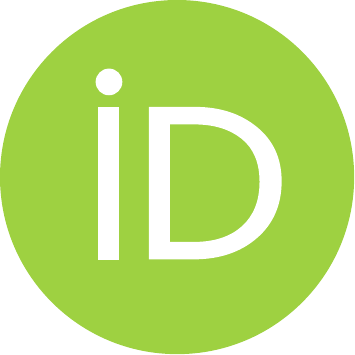}}}%
  \thanks{%
    Manuscript received January 19, 2015; revised April 2, 2015 and September 11, 2015; accepted October 15, 2015.\newline
    Copyright \copyright\ 2015 IEEE. Personal use of this material is permitted. However, permission to use this material for any other purposes must be obtained from the IEEE by sending a request to \protect\url{pubs-permissions@ieee.org}.\newline
    G.~Herbst is with Siemens AG, Digital Factory Division, 09116 Chemnitz, Germany (e-mail: \protect\url{gernot.herbst@siemens.com}).%
  }
}

\markboth%
{IEEE Transactions on Industrial Electronics}
{Herbst: Practical Active Disturbance Rejection Control: Bumpless Transfer, Rate Limitation and Incremental Algorithm}

\maketitle

\begin{abstract}
Practical applications of controllers often impose further requirements on the implementation beyond the actual control performance, such as the ability to switch between manual and automatic control or between different control laws or controller parameter settings, known as bumpless transfer. Another common requirement is to limit the control signal in magnitude and/or rate. This article examines and extends several discrete-time variants of active disturbance rejection control (ADRC), which is increasingly being applied especially in the field of power electronics and drives, in this regard. Detailed guidelines for practical ADRC implementations with these abilities are presented, and all features are being demonstrated with the help of simulation examples.
\end{abstract}

\begin{IEEEkeywords}
Active disturbance rejection control (ADRC), incremental algorithm, bumpless transfer.
\end{IEEEkeywords}

\IEEEpeerreviewmaketitle


\section{Introduction}

\IEEEPARstart{A}{ctive} disturbance rejection control (ADRC) has arisen as a control paradigm that adopts a broader sense of disturbances in control loops \cite{Han:2009,Gao:2006,Gao:2014}. While incorporating elements from modern control theory, the user is being relieved from the necessity of detailed plant modelling, and tuning the controller becomes quite straightforward \cite{Chen:2011}.

Apart from extensions to the ADRC concept such as handling of time-delay systems \cite{Zhao:2014} and theoretical studies such as the examination of its stability properties \cite{Zheng:PhD2009}, a significant increase in applications of ADRC across different domains can be observed in recent years \cite{Zheng:2010}, ranging from power electronics \cite{Sun:2005,Xu:2015} to machines and motion control \cite{Su:2005,Sira:2014,Wu:2013,Li:2015} and other areas \cite{Tang:2014,Xue:2015,Zhao:2015}. With the rise of practical applications of ADRC, it becomes increasingly necessary to cover several practical aspects of an ADRC implementation that are often overlooked in theory-focused studies. To that end, this article will develop guidelines for a discrete-time implementation (both non-incremental and, as a new addition, incremental) that enables a magnitude and/or rate limitation of the control signal as well as bumpless changes to the controller.

The remainder of this article is organised as follows: Since a digital implementation of ADRC is being assumed for almost any practical application, the discretisation and parameterisation of linear ADRC will be repeated in \refSec{sec:adrc}. Based upon that, an incremental formulation of ADRC is being introduced in \refSec{sec:incremental}, which can be employed as a replacement or in conjunction with other incremental controller algorithms. As the main contribution of this article, \refSec{sec:limitation} and \ref{sec:bumpless} are devoted to two major aspects important to a practical ADRC implementation: magnitude and rate limitation of the control signal generated by an ADRC controller on the one hand, and performing bumpless changes to the controller (online parameter changes and enabling the controller) on the other hand. As a result of this article, all steps are described that are required for a non-incremental or incremental ADRC implementation with or without magnitude/rate limitation and the ability for bumpless changes to the controller.


\section{Review of Discrete-Time ADRC}
\label{sec:adrc}


\subsection{Brief Summary of Continuous-Time ADRC}
\label{sec:adrc_continuous}

We start with a simple model of a linear $n$-th order plant with output $y$, input $u$ and input disturbance $d$ as given in \refEq{eqn:ptn}:
\begin{equation}
y^{(n)}(t) + \sum_{i=0}^{n-1} a_i \cdot y^{(i)}(t) = b \cdot u(t) + e \cdot d(t)
\label{eqn:ptn}
\end{equation}

In \refEq{eqn:ptn}, $b$ is now being split into a (known) approximate value $b_0$ and an unknown remainder $\Delta b$. For the controller design, $b_0$ will be everything one needs to know about a plant when using ADRC. With $b = b_0 + \Delta b$ we rearrange \refEq{eqn:ptn} into \refEq{eqn:ptn2}. The key idea of ADRC is to assume an $n$-th order integrator behaviour for any $n$-th order model and to combine both actual disturbances and modelling errors (including all derivatives $y^{(i)}(t)$ for $i < n$, which are being ignored in the model) in a so-called generalised disturbance $f$:
\begin{align}
y^{(n)}(t) &= \underbrace{ -\sum_{i=0}^{n-1} a_i \cdot y^{(i)}(t) + e \cdot d(t) + \Delta b \cdot u(t) }_{\text{generalised disturbance}\, f(t)} + b_0 \cdot u(t)  \notag\\
&= f(t) + b_0 \cdot u(t)
\label{eqn:ptn2}
\end{align}

In many practical applications, plants are dominated by a first- or second-order low-pass behaviour. For these two important cases, the modelling approach and generalised disturbance will be given in the following:
\begin{enumerate}
\item
$T \cdot \dot{y}(t) + y(t) = K \cdot u(t) + d(t)$

\smallskip
is being modelled as a first-order integrator as follows:
\smallskip

$
\begin{aligned}
\dot{y}(t)
&= -\textstyle\frac{1}{T} \cdot y(t) + \frac{1}{T} \cdot d(t) + \Delta b \cdot u(t) + b_0 \cdot u(t)  \\
&= f(t) + b_0 \cdot u(t)
\quad \text{with} \quad b_0 \approx \textstyle\frac{K}{T}
\end{aligned}
$
\medskip

\item
$T^2 \cdot \ddot{y}(t) + 2DT \cdot \dot{y}(t) + y(t) = K \cdot u(t) + d(t)$

\smallskip
is being modelled as a second-order integrator:
\smallskip

$
\begin{aligned}
\ddot{y}(t)
&= -\textstyle\frac{2D}{T} \cdot \dot{y}(t) - \frac{1}{T^2} \cdot y(t) + \frac{1}{T^2} \cdot d(t) + \Delta b \cdot u(t)  \\
&\phantom{=\ } + b_0 \cdot u(t)  \notag\\
&= f(t) + b_0 \cdot u(t)
\quad \text{with} \quad b_0 \approx \textstyle\frac{K}{T^2}
\end{aligned}
$
\medskip

\end{enumerate}

In linear ADRC, a \Person{Luenberger} observer is then being set up, where the process model includes a constant input disturbance, which will be the estimation of the generalised disturbance $f$:
\begin{equation}
\Vector{\dot{\hat{x}}}(t)
= (\Matrix{A} - \Matrix{L}\Matrix{C}) \cdot \Vector{\hat{x}}(t) + \Matrix{B} \cdot \Vector{u}(t) +  \Matrix{L} \cdot y(t)
\label{eqn:adrc_observer}
\end{equation}

For the first- and second-order case the respective matrices $\Matrix{A}$, $\Matrix{B}$ and $\Matrix{C}$ are:
\begin{enumerate}
\item
$\Matrix{A} =
\begin{pmatrix}
0 & 1 \\
0 & 0
\end{pmatrix}$,\quad
$\Matrix{B} =
\begin{pmatrix}
b_0 \\ 0
\end{pmatrix}$,\quad
$\Matrix{C} =
\begin{pmatrix}
1 & 0
\end{pmatrix}$
\smallskip

\item
$\Matrix{A} =
\begin{pmatrix}
0 & 1 & 0 \\
0 & 0 & 1 \\
0 & 0 & 0 \\
\end{pmatrix}$,\quad
$\Matrix{B} =
\begin{pmatrix}
0 \\ b_0 \\ 0
\end{pmatrix}$,\quad
$\Matrix{C} =
\begin{pmatrix}
1 & 0 & 0
\end{pmatrix}$
\end{enumerate}

Using the estimated state variables $\Vector{\hat{x}}$, a state space controller with disturbance rejection is employed \cite{Herbst:2013}, which can in general be formulated as follows:
\begin{equation}
u(t) = \D\frac{K_\mathrm{P}}{b_0} \cdot r(t) - \Vector{w}^\Transpose \cdot \Vector{\hat{x}}(t)
\label{eqn:adrc_standard_u}
\end{equation}

\begin{figure}
    \centering%
    \includegraphics[width=\linewidth]{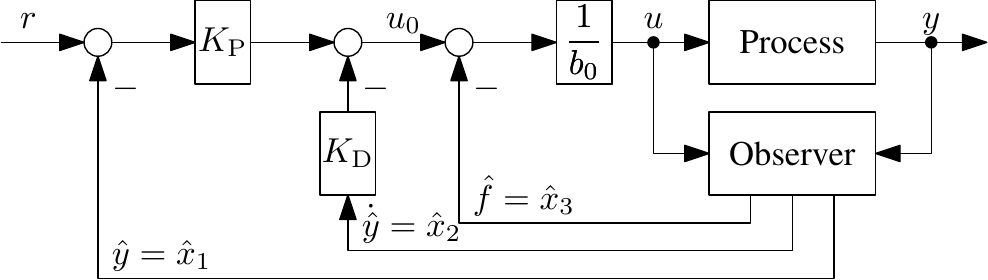}%
    \caption{Control loop structure with active disturbance rejection control (ADRC) for a second-order process.}
    \label{fig:adrc2_structure}
\end{figure}

For the second-order case, the structure of the control loop is given in \refFig{fig:adrc2_structure}. In \refEq{eqn:adrc_standard_u}, the feedback gain vector $\Vector{w}$ is chosen depending on the order of the dominant plant dynamics:
\begin{equation}
\Vector{w}^\Transpose =
\begin{cases}
\begin{pmatrix} K_\mathrm{P}  &  1  \end{pmatrix} \cdot \frac{1}{b_0}  &  \text{(first-order case)}  \\
\begin{pmatrix} K_\mathrm{P}  &  K_\mathrm{D}  &  1  \end{pmatrix} \cdot \frac{1}{b_0}  &  \text{(second-order case)}  \\
\end{cases}
\label{eqn:adrc_standard_w}
\end{equation}

For the first- and second-order case, these controllers resemble P- and PD-controllers acting on first- and second-order integrator plants $y^{(1/2)}(t) = f(t) + b_0 \cdot u(t)$, respectively. Tuning $K_\mathrm{P}$ and $K_\mathrm{D}$ in \refEq{eqn:adrc_standard_w} according to the desired closed-loop dynamics is therefore straightforward.  As a final step, the observer gains $\Matrix{L}$ in \refEq{eqn:adrc_observer} have to be chosen such that the observer poles are far enough left of the closed-loop poles in order to enable the disturbance observer to handle both actual disturbances and the (intentional) modelling error.

A common choice is to place all closed-loop poles at one location $s^\mathrm{CL} = -\omega^\mathrm{CL}$ (bandwidth parameterisation, cf.\ \cite{Gao:2003}) and all poles of the extended state observer at $s^\mathrm{ESO} \approx (3 \ldots 10) \cdot s^\mathrm{CL}$.

This concludes the brief summary of the continuous-time case of linear ADRC. A more detailed and elaborate introduction can be found in \cite{Herbst:2013}.


\subsection{Discretisation}
\label{sec:adrc_discrete}

For a discrete-time implementation, the matrices $\Matrix{A}$, $\Matrix{B}$ and $\Matrix{C}$ from the the state space process models can be transformed into their discrete-time counterparts by zero-order hold (ZOH) discretisation as follows:
\begin{equation}
\begin{aligned}
\Matrix{A}_\mathrm{d} &= \Matrix{I} + \sum_{i = 1}^\infty \D\frac{\Matrix{A}^i \cdot T^i_\mathrm{sample}}{i!}  \\
\Matrix{B}_\mathrm{d} &= \left( \sum_{i = 1}^\infty \D\frac{\Matrix{A}^{i-1} \cdot T^i_\mathrm{sample}}{i!} \right) \cdot \Matrix{B}  \\
\Matrix{C}_\mathrm{d} &= \Matrix{C}  \\
\end{aligned}
\label{eqn:zoh_transform}
\end{equation}

\begin{figure}
    \centering%
    \includegraphics{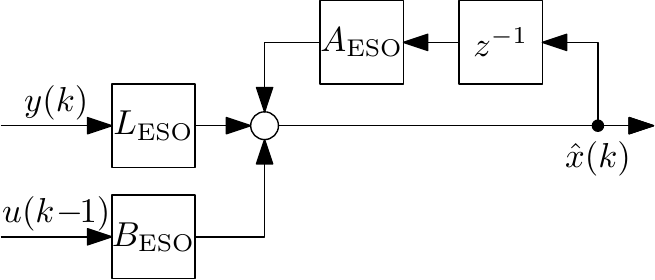}%
    \caption{Discrete-time extended state observer as given in \refEq{eqn:adrc_standard_x}.}
    \label{fig:adrc_standard_eso}
\end{figure}

For the discrete-time observer in ADRC, it is beneficial to employ the current observer (depicted in \refFig{fig:adrc_standard_eso}) as opposed to the predictive observer approach in order to include the most recent measurement $y(k)$ for the estimation of $\Vector{\hat{x}}(k)$, cf.\ \cite{Miklosovic:2006}:
\begin{equation}
\Vector{\hat{x}}(k) = \Matrix{A}_\mathrm{ESO} \cdot \Vector{\hat{x}}(k-1) + \Matrix{B}_\mathrm{ESO} \cdot u(k-1) + \Matrix{L}_\mathrm{ESO} \cdot y(k)
\label{eqn:adrc_standard_x}
\end{equation}
with
\begin{equation}
\begin{aligned}
\Matrix{A}_\mathrm{ESO} &= \Matrix{A}_\mathrm{d} - \Matrix{L}_\mathrm{c} \cdot \Matrix{C}_\mathrm{d} \cdot \Matrix{A}_\mathrm{d}  \\
\Matrix{B}_\mathrm{ESO} &= \Matrix{B}_\mathrm{d} - \Matrix{L}_\mathrm{c} \cdot \Matrix{C}_\mathrm{d} \cdot \Matrix{B}_\mathrm{d}  \\
\Matrix{L}_\mathrm{ESO} &= \Matrix{L}_\mathrm{c}  \\
\end{aligned}
\label{eqn:adrc_discrete_ABLESO}
\end{equation}

\begin{figure}
    \centering%
    \includegraphics{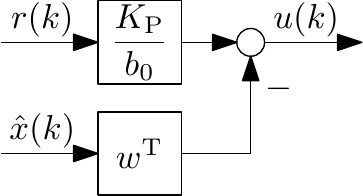}%
    \caption{Discrete-time control law as given in \refEq{eqn:adrc_discrete_u}.}
    \label{fig:adrc_standard_controller}
\end{figure}

The discrete-time control law (depicted in \refFig{fig:adrc_standard_controller}) does not change (apart from exchanging time $t$ with the current sample point $k$), and the same feedback gain vector $\Vector{w}$ as given in \refEq{eqn:adrc_standard_w} is being used:
\begin{equation}
u(k) = \D\frac{K_\mathrm{P}}{b_0} \cdot r(k) - \Vector{w}^\Transpose \cdot \Vector{\hat{x}}(k)
\label{eqn:adrc_discrete_u}
\end{equation}

\subsection{Parameterisation}
\label{sec:adrc_parameter}

For the control law \refEq{eqn:adrc_discrete_u}, one needs to choose $b_0$, $K_\mathrm{P}$ and (only for second-order ADRC) $K_\mathrm{D}$. While $b_0$ represents the approximate knowledge of the process (cf.\ \refSec{sec:adrc_continuous}), the actual controller parameters $K_\mathrm{P}$ and $K_\mathrm{D}$ can, for instance, be set by placing the closed-loop poles according to a desired settling time. For first- and second-order ADRC, this leads to:
\begin{enumerate}
\item
$K_\mathrm{P} = -s^\mathrm{CL}$
\ with \
$s^\mathrm{CL} \approx -\D\frac{4}{T_\mathrm{settle}}$

\item
$
K_\mathrm{P} = \left(s^\mathrm{CL}\right)^2$%
,\
$K_\mathrm{D} = -2 \cdot s^\mathrm{CL}$
\ with \
$s^\mathrm{CL} \approx -\D\frac{6}{T_\mathrm{settle}}
$
\end{enumerate}

The observer poles have to be placed far enough to the left of the closed-loop poles in the $s$-plane, e.\,g.\ by a common location $s^\mathrm{ESO} = k_\mathrm{ESO} \cdot s^\mathrm{CL} \approx (3 \ldots 10) \cdot s^\mathrm{CL}$. However a compromise between disturbance rejection and noise sensitivity has to be found when placing the observer poles \cite{Herbst:2013}. For the discrete-time implementation, the desired observer poles have to be mapped to the $z$-plane via $z^\mathrm{ESO} = \mathrm{e}^{s^\mathrm{ESO} \cdot T_\mathrm{sample}}$. Finally the necessary observer gains can be computed \cite{Miklosovic:2006}:
\begin{equation}
\Matrix{L}_\mathrm{c}
=
\begin{cases}
\begin{pmatrix}
1 - \left(z^\mathrm{ESO}\right)^2  \\[0.5ex]
\frac{\left(1 - z^\mathrm{ESO}\right)^2}{T_\mathrm{sample}}  \\
\end{pmatrix}
&  \text{(first-order ADRC)} \\[4ex]
\begin{pmatrix}
1 - \left(z^\mathrm{ESO}\right)^3  \\[0.5ex]
\frac{3 \cdot \left(1 - z^\mathrm{ESO}\right)^2 \cdot \left(1 + z^\mathrm{ESO}\right)}{2 \cdot T_\mathrm{sample}}  \\[1ex]
\frac{\left(1 - z^\mathrm{ESO}\right)^3}{T^2_\mathrm{sample}}  \\
\end{pmatrix}
&  \text{(second-order ADRC)}  \\
\end{cases}
\label{eqn:adrc_discrete_Lc}
\end{equation}


\subsection{Lag-Reduced Form and Implementation}
\label{sec:adrc_opt}

\begin{figure}
    \centering%
    \includegraphics{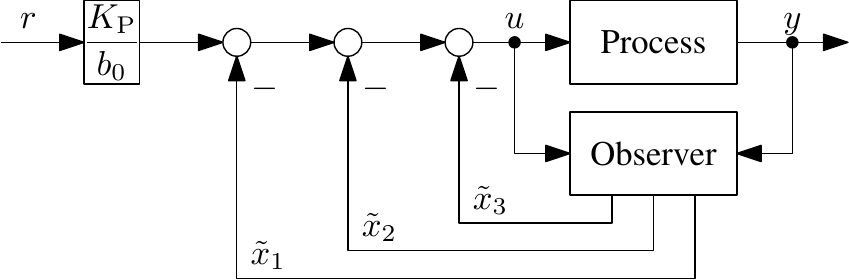}%
    \caption{Control loop with modified (second-order) ADRC structure and transformed state variables $\tilde{x}$ according to \refEq{eqn:adrc_opt_x} and \refEq{eqn:adrc_opt_u}.}
    \label{fig:adrc_structure_modified}
\end{figure}

For time-critical applications, the discrete-time implementation of ADRC can be improved regarding its computational lag to a certain extent, especially the computation of the next control signal value $u(k)$ after a new measurement $y(k)$ has become available. In \cite{Herbst:2013}, an idea was presented to reduce the computational effort in the control law \refEq{eqn:adrc_discrete_u} by using a transformed state vector $\Vector{\tilde{x}} = \Matrix{T}^{-1} \cdot \Vector{\hat{x}}$ such that the feedback gain vector $\Vector{w}$ is reduced to an all-ones vector matching the number of observed state variables, cf.\ \refFig{fig:adrc_structure_modified}. The control law then becomes:
\begin{equation}
u(k) = \D\frac{K_\mathrm{P}}{b_0} \cdot r(k) - \Vector{w}^\Transpose \cdot \Vector{\tilde{x}}(k)
\label{eqn:adrc_opt_u}
\end{equation}
with
\begin{equation}
\Vector{w}^\Transpose =
\begin{cases}
\begin{pmatrix} 1  &  1  \end{pmatrix}  &  \text{(first-order ADRC)}  \\
\begin{pmatrix} 1  &  1  &  1  \end{pmatrix}  &  \text{(second-order ADRC)}  \\
\end{cases}
\label{eqn:adrc_opt_w}
\end{equation}

The observer has to be adapted accordingly:
\begin{equation}
\Vector{\tilde{x}}(k) = \Matrix{\tilde{A}}_\mathrm{ESO} \cdot \Vector{\tilde{x}}(k-1) + \Matrix{\tilde{B}}_\mathrm{ESO} \cdot u(k-1) + \Matrix{\tilde{L}}_\mathrm{ESO} \cdot y(k)
\label{eqn:adrc_opt_x}
\end{equation}
with
\begin{equation}
\begin{aligned}
\Matrix{\tilde{A}}_\mathrm{ESO} &= \Matrix{T}^{-1} \cdot \Matrix{A}_\mathrm{ESO} \cdot \Matrix{T}  \\
\Matrix{\tilde{B}}_\mathrm{ESO} &= \Matrix{T}^{-1} \cdot \Matrix{B}_\mathrm{ESO}  \\
\Matrix{\tilde{L}}_\mathrm{ESO} &= \Matrix{T}^{-1} \cdot \Matrix{L}_\mathrm{ESO}  \\
\end{aligned}
\label{eqn:adrc_opt_ABLESO}
\end{equation}

The transformation matrix $\Matrix{T}^{-1}$ needed to obtain the transformed state variables leading to the simplified feedback gain vector is:
\begin{equation}
\Matrix{T}^{-1}
=
\begin{cases}
\D\frac{1}{b_0} \cdot
\begin{pmatrix}
K_\mathrm{P} & 0 \\
0 & 1 \\
\end{pmatrix}
&  \text{(first-order ADRC)} \\
\D\frac{1}{b_0} \cdot
\begin{pmatrix}
K_\mathrm{P} & 0 & 0 \\
0 & K_\mathrm{D} & 0 \\
0 & 0 & 1 \\
\end{pmatrix}
&  \text{(second-order ADRC)}  \\
\end{cases}
\label{eqn:adrc_opt_tinv}
\end{equation}

The computation of the observer and controller equations can be simplified further such that an updated control signal value $u(k)$ can be provided requiring only one addition and one multiplication after a new measurement $y(k)$ has become available, cf.\ \cite{Herbst:2013} for details.


\section{Incremental ADRC}
\label{sec:incremental}

As a first contribution of this article, a novel incremental form of ADRC will be introduced in this section, resulting in a control law for an incremental control signal value $\Delta u(k)$.

In order to derive a set of incremental equations, one has to substitute \refEq{eqn:adrc_inc_standard_x} into \refEq{eqn:adrc_standard_x} and solve for $\Delta\Vector{\hat{x}}$. For the incremental control signal value $\Delta u$, \refEq{eqn:adrc_inc_standard_u} has to be put in \refEq{eqn:adrc_discrete_u}. The incremental implementation of ADRC consists of the following four equations:
\begin{figure}
    \centering%
    \includegraphics{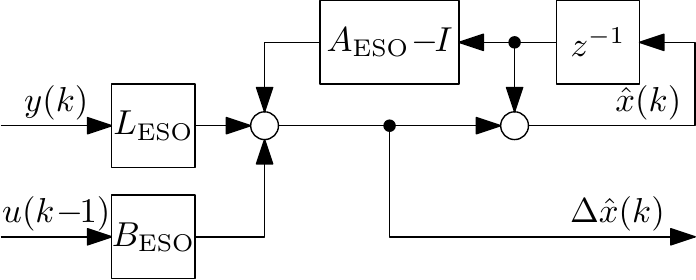}%
    \caption{Discrete-time extended state observer with incremental output as given in \refEq{eqn:adrc_inc_standard_deltax}.}
    \label{fig:adrc_inc_eso}
\end{figure}
\begin{equation}
\Delta\Vector{\hat{x}}(k) = \left( \Matrix{A}_\mathrm{ESO} - \Matrix{I} \right) \cdot \Vector{\hat{x}}(k-1) + \Matrix{B}_\mathrm{ESO} \cdot u(k-1) + \Matrix{L}_\mathrm{ESO} \cdot y(k)
\label{eqn:adrc_inc_standard_deltax}
\end{equation}
\begin{equation}
\Vector{\hat{x}}(k) = \Vector{\hat{x}}(k-1) + \Delta\Vector{\hat{x}}(k)
\label{eqn:adrc_inc_standard_x}
\end{equation}
\begin{equation}
\Delta u(k) = \D\frac{K_\mathrm{P}}{b_0} \cdot \Delta r(k) - \Vector{w}^\Transpose \cdot \Delta\Vector{\hat{x}}(k)
\label{eqn:adrc_inc_standard_deltau}
\end{equation}
\begin{equation}
u(k) = u(k-1) + \Delta u(k)
\label{eqn:adrc_inc_standard_u}
\end{equation}

The incremental observer as given in \refEq{eqn:adrc_inc_standard_deltax} and \refEq{eqn:adrc_inc_standard_x} is depicted in \refFig{fig:adrc_inc_eso}. If the incremental implementation shall be based on the lag-reduced form from \refSec{sec:adrc_opt}, the structure of these equations does not change, only the matrices $\Matrix{A}_\mathrm{ESO}$, $\Matrix{B}_\mathrm{ESO}$ and $\Matrix{L}_\mathrm{ESO}$ as well as the state vector $\Vector{\hat{x}}$ have to be exchanged with their transformed counterparts $\Matrix{\tilde{A}}_\mathrm{ESO}$, $\Matrix{\tilde{B}}_\mathrm{ESO}$, $\Matrix{\tilde{L}}_\mathrm{ESO}$ and $\Vector{\tilde{x}}$, respectively. Additionally, the feedback gain vector $\Vector{w}$ from \refEq{eqn:adrc_opt_w} has to be used in \refEq{eqn:adrc_inc_standard_deltau} in this case.

Regarding their functionality and control performance, the non-incremental and incremental forms of ADRC are identical. However, a rate limitation of the control signal can be implemented in a more straightforward manner when using the incremental form. It also facilitates easier switching between different incremental controller types that act on the same integrator, or when using actuators with integral behaviour.


\section{Magnitude and Rate Limitation}
\label{sec:limitation}

\begin{figure}
    \centering%
    \includegraphics{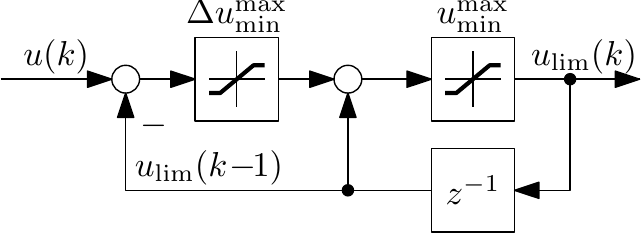}%
    \caption{Discrete-time magnitude and rate limiter.}
    \label{fig:limiter}
\end{figure}

A typical problem that can arise if control signals run into saturation is integral windup of the controller. The second contribution of this article, presented in the following, is therefore to examine the non-incremental and incremental forms of ADRC in this regard. As will be demonstrated, even a rate limitation of the control signal can be implemented relatively easily, which has to be seen as a major practical feature of ADRC for industrial applications.

Regarding the control signal saturation problem, \cite{Herbst:2013} indicates that ADRC does not suffer from integral windup if the saturated control signal is fed into the observer. This is a technique already known from \cite{Astrom:1989} in order to prevent windup. The same holds true if a control signal is rate-limited, e.\,g.\ when using a magnitude and rate limiter as depicted in \refFig{fig:limiter}, which is described as follows:
\begin{equation}
\begin{aligned}
u_\mathrm{lim}(k) =
\limit_{u_\mathrm{min}}^{u_\mathrm{max}} \Big( & u_\mathrm{lim}(k-1) + \\
  &  \limit_{\Delta u_\mathrm{min}}^{\Delta u_\mathrm{max}} \big( u(k)- u_\mathrm{lim}(k-1) \big) \Big)
\end{aligned}
\end{equation}
with
\begin{equation}
\mylimit{x}{x_\mathrm{min}}{x_\mathrm{max}} =
\begin{cases}
x_\mathrm{min}  & ,\ x < x_\mathrm{min} \\
x  & ,\ x_\mathrm{min} \le x \le x_\mathrm{max}  \\
x_\mathrm{max}  & ,\ x > x_\mathrm{max}  \\
\end{cases}
\end{equation}

\begin{figure}
    \centering%
    \includegraphics[width=\linewidth]{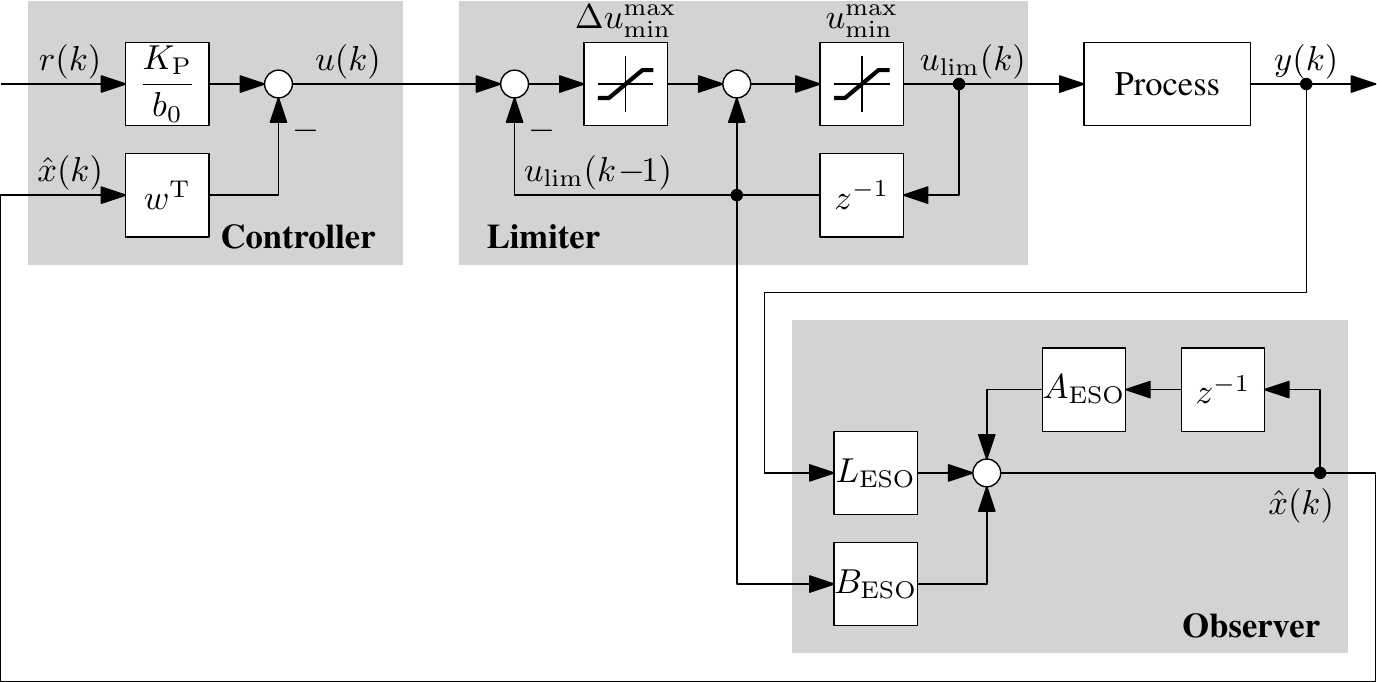}%
    \caption{Structure and implementation of a control loop based on discrete-time ADRC with magnitude and rate limitation.}
    \label{fig:adrc_limited}
\end{figure}

For the observer in \refFig{fig:adrc_standard_eso} this means that the limited control signal value $u_\mathrm{lim}(k-1)$ has to be used instead of $u(k-1)$. In \refFig{fig:adrc_limited}, the structure of a complete control loop with discrete-time ADRC and its components controller, observer and limiter is shown.

\begin{figure}
    \centering%
    \includegraphics{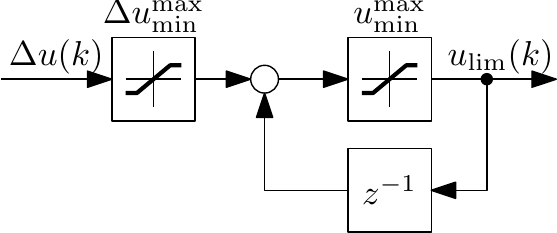}%
    \caption{Discrete-time magnitude and rate limiting integrator.}
    \label{fig:limiter_inc}
\end{figure}

For the incremental implementation of ADRC introduced in \refSec{sec:incremental}, however, additional measures have to be taken to maintain the controller performance while allowing a magnitude- and/or rate-limited control signal, e.\,g.\ when using a limiting integrator such as \refFig{fig:limiter_inc} instead of \refEq{eqn:adrc_inc_standard_u}. More precisely, one must ensure that the stationary value of the control signal $u$ necessary for compensating the control error can be reached even when being reconstructed from a summation of limited $\Delta u$-values that will differ from the unlimited values computed by \refEq{eqn:adrc_inc_standard_deltau}.

In other words, the desired control signal value computed before the limitation is being applied should not be forgotten. To ensure this, \refEq{eqn:adrc_inc_standard_deltau} is extended by a new carry-over term that represents the difference between the desired value of $\Delta u$ at time $k-1$ and the actual difference value of the magnitude- and/or rate-limited control signal $\Delta u_\mathrm{lim}(k-1)$:
\begin{equation}
\begin{split}
\Delta u(k) =
& \D\frac{K_\mathrm{P}}{b_0} \cdot \Delta r(k) - \Vector{w}^\Transpose \cdot \Delta\Vector{\hat{x}}(k) \\
&+ \left( \Delta u(k-1) - \Delta u_\mathrm{lim}(k-1) \right) \\
\end{split}
\label{eqn:adrc_limited_deltau}
\end{equation}
with
\begin{equation}
\Delta u_\mathrm{lim}(k-1) = u_\mathrm{lim}(k-1) - u_\mathrm{lim}(k-2)
\end{equation}

\begin{figure}
    \centering%
    \includegraphics{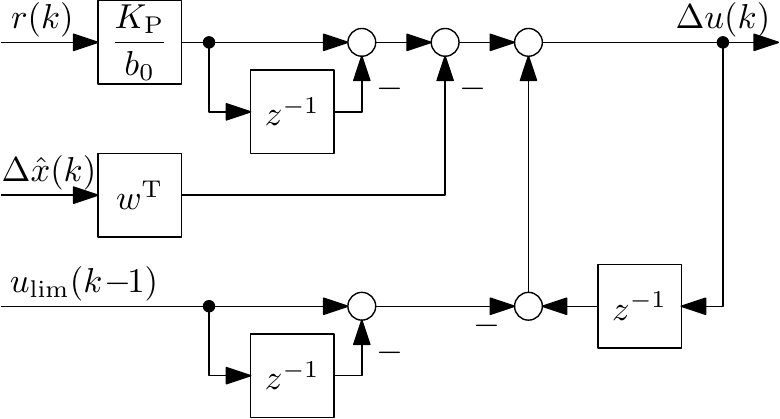}%
    \caption{Discrete-time control law \refEq{eqn:adrc_limited_deltau} with incremental output and feedback of the magnitude- and/or rate-limited control signal.}
    \label{fig:adrc_controller_incremental}
\end{figure}

In \refFig{fig:adrc_controller_incremental} the modified control law \refEq{eqn:adrc_limited_deltau} for incremental ADRC with limiter case is shown. In \refEq{eqn:adrc_limited_deltau}, the magnitude and/or rate limitation is being modelled as follows:
\begin{equation}
u_\mathrm{lim}(k) =
\mylimit{
  u_\mathrm{lim}(k-1) + \mylimit{\Delta u(k)}{\Delta u_\mathrm{min}}{\Delta u_\mathrm{max}}
}{u_\mathrm{min}}{u_\mathrm{max}}
\end{equation}

\begin{figure}
    \centering%
    \includegraphics[width=\linewidth]{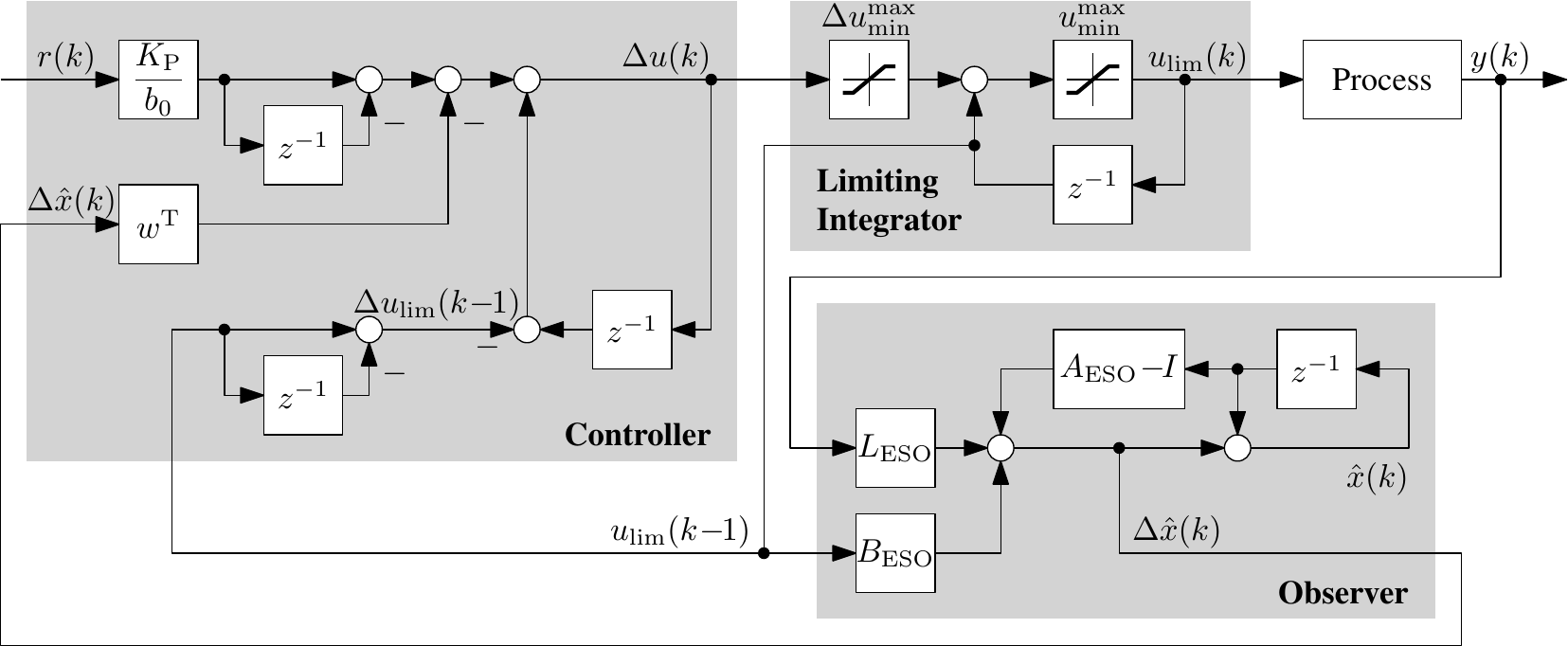}%
    \caption{Structure and implementation of a control loop based on discrete-time incremental ADRC with magnitude and rate limitation.}
    \label{fig:adrc_limited_incremental}
\end{figure}

As a summary, the complete structure of a control loop with discrete-time incremental ADRC and a magnitude/rate-limited control signal is shown in \refFig{fig:adrc_limited_incremental}.


\section{Bumpless Transfer}
\label{sec:bumpless}

Apart from the actual controller performance, there are other important features that should be present in a robust practical implementation of a controller, which can be subsumed as bumpless transfer: Jumps in the control signal should be avoided when enabling a controller (i\,.e.\ switching from manual mode to automatic control), when switching between different controllers, or when changing parameters while the controller is enabled.

As the third novel contribution of this article, the behaviour of discrete-time ADRC will be examined in the following with regard to these features. The necessary measures in order to achieve bumpless transfer will be presented for all flavours of discrete-time ADRC presented in this article, i.\,e.\ the four possible combinations of incremental vs.\ non-incremental implementation (cf.\ \refSec{sec:incremental}) and lag-reduced vs.\ standard form of the observer and state variables (cf.\ \refSec{sec:adrc_opt}).

The guidelines for bumpless transfer presented in this section assume that changes are performed in a stationary state while $r = y$ holds, but can also be extended for changes during transients. One main principle is to adjust the estimated disturbance in order to keep the control signal value constant.


\subsection{Enabling the ESO}
\label{sec:bumpless_enable_eso}

Since ADRC is built on an observer, the control law cannot be executed before the observer is enabled. For all variants of discrete-time ADRC presented in this paper, the observer can be enabled simply by executing the respective update equation of the observer for the first time.

Preferably, the observer should be enabled some time before the controller computes its first control signal value, such that the estimation can reach a steady state. To speed up this procedure or when the observer may be enabled only together with the actual controller, the initial value of the state vector $\Vector{\hat{x}}(k-1)$ may be initialised with the process output $y(k-1)$ and the control signal $u(k-1)$, for example in the second-order case:
\begin{equation}
\Vector{\hat{x}}(k-1) = \begin{pmatrix} y(k-1) &  0  &  -b_0 \cdot u(k-1)  \end{pmatrix}^\Transpose
\label{eqn:eso_init_x}
\end{equation}

If the lag-reduced form of the state variables is being used, the transformation must be applied to the initial value:
\begin{align}
\Vector{\tilde{x}}(k-1) &= \Matrix{T}^{-1} \cdot \begin{pmatrix} y(k-1) &  0  &  -b_0 \cdot u(k-1)  \end{pmatrix}^\Transpose
\notag\\
&= \begin{pmatrix} \frac{K_\mathrm{P}}{b_0} \cdot y(k-1) &  0  &  -u(k-1)  \end{pmatrix}^\Transpose
\label{eqn:eso_init_xtilde}
\end{align}

Both \refEq{eqn:eso_init_x} and \refEq{eqn:eso_init_xtilde} assume to be used at time $k$ directly before executing the update equation of the observer (i.\,e.\ \refEq{eqn:adrc_standard_x}, \refEq{eqn:adrc_opt_x} or \refEq{eqn:adrc_inc_standard_deltax} depending on the ADRC variant) for the first time.


\subsection{Enabling the Controller}
\label{sec:bumpless_enable_controller}

For the non-incremental variants of ADRC, the controller can be enabled at any time, provided the observer has been enabled before or, at the latest, jointly with the control law. For the incremental variants, however, the observer should be enabled before the controller and the value $\Delta u(k-1)$ (cf.\ \refFig{fig:adrc_controller_incremental}) has to be set up with an initial value as follows:
\begin{equation}
\Delta u(k-1) = \frac{K_\mathrm{P}}{b_0} \cdot r(k-1) - \Vector{w}^\Transpose \cdot \Vector{\hat{x}}(k-1) - u_\mathrm{lim}(k-2)
\label{eqn:controller_init_deltau}
\end{equation}

From \refEq{eqn:controller_init_deltau} it becomes apparent that the values $r(k-1)$, $\Vector{\hat{x}}(k-1)$ and $u_\mathrm{lim}(k-2)$ must be made available at time $k$. $u_\mathrm{lim}(k-2)$ is the actual (limited) value of the control action at time $k-2$, i.\,e.\ before enabling the ADRC control law. It may either contain a manually set value (when switching from manual mode to ADRC) or stem from other controllers (when switching between different control laws).

\refEq{eqn:controller_init_deltau} applies to incremental ADRC with transformed state variables according to \refSec{sec:adrc_opt} as well, only \refEq{eqn:adrc_opt_w} has to be applied for $\Vector{w}^\Transpose$ instead of \refEq{eqn:adrc_standard_w} when using transformed state variables $\Vector{\tilde{x}}$ instead of $\Vector{\hat{x}}$.

Similar to \refSec{sec:bumpless_enable_eso}, we assume that the initialisation with \refEq{eqn:controller_init_deltau} is performed at time $k$ directly before computing the first value of the control action.


\subsection{Parameter Changes}
\label{sec:bumpless_parameter}

In this section, changes to the ADRC parameters will be treated in three separate groups:
\begin{itemize}
\item
changes to the plant model, i.\,e.\ changing $b_0$,

\item
changes to the observer, i.\,e.\ moving the observer poles, and

\item
changes to the control law, i.\,e.\ moving the closed-loop poles by changing $K_\mathrm{P}$ and/or $K_\mathrm{D}$.
\end{itemize}

In the following, values of parameters and variables prior to the parameter change will be denoted by an ``old'' superscript. Also, as in \refSec{sec:bumpless_enable_eso} and \refSec{sec:bumpless_enable_controller}, it is assumed that parameter changes are performed at time $k$, and the necessary computations (if any) as described in the following are carried out directly before running the observer and controller equations for time $k$.

\subsubsection{Moving the closed-loop poles}

For the ADRC variants with transformed state variables, the transformation matrix $\Matrix{T}^{-1}$ has to be populated with the new values of $K_\mathrm{P}$ and/or $K_\mathrm{D}$, cf.\ \refEq{eqn:adrc_opt_tinv}. Due to this change, the observer matrices and the state variables have to be retransformed as follows:
\begin{equation}
\begin{aligned}
\Matrix{\tilde{A}}_\mathrm{ESO} &= \left( \Matrix{T}^{-1} \cdot \Matrix{T}^\mathrm{old} \right) \cdot \Matrix{A}_\mathrm{ESO}^\mathrm{old} \cdot \left( \Matrix{T}^{-1} \cdot \Matrix{T}^\mathrm{old} \right)^{-1}  \\
\Matrix{\tilde{B}}_\mathrm{ESO} &= \left( \Matrix{T}^{-1} \cdot \Matrix{T}^\mathrm{old} \right) \cdot \Matrix{B}_\mathrm{ESO}^\mathrm{old}  \\
\Matrix{\tilde{L}}_\mathrm{ESO} &= \left( \Matrix{T}^{-1} \cdot \Matrix{T}^\mathrm{old} \right) \cdot \Matrix{L}_\mathrm{ESO}^\mathrm{old}  \\
\Vector{\tilde{x}}(k-1) &= \left( \Matrix{T}^{-1} \cdot \Matrix{T}^\mathrm{old} \right) \cdot \Vector{\tilde{x}}^\mathrm{old}(k-1)  \\
\end{aligned}
\label{eqn:change_parameter_controller1}
\end{equation}

With a small simplification, one can express $\Matrix{T}^{-1} \cdot \Matrix{T}^\mathrm{old}$ e.\,g.\ for the second-order case as:
\begin{equation*}
\Matrix{T}^{-1} \cdot \Matrix{T}^\mathrm{old}
=
\begin{pmatrix}
K_\mathrm{P} / K_\mathrm{P}^\mathrm{old}  &  0  &  0  \\
0  &  K_\mathrm{D} / K_\mathrm{D}^\mathrm{old}  &  0  \\
0  &  0  &  1 \\
\end{pmatrix}
\end{equation*}

For incremental ADRC variants, $\Delta u(k-1)$ has to be recomputed additionally, again using \refEq{eqn:controller_init_deltau}.

\subsubsection{Moving the observer poles}

When choosing a new common location for the observer poles, the observer gain matrix $\Matrix{L}_\mathrm{ESO} = \Matrix{L}_\mathrm{c}$ has to be updated according to $\refEq{eqn:adrc_discrete_Lc}$, followed by updating $\Matrix{A}_\mathrm{ESO}$ and $\Matrix{B}_\mathrm{ESO}$ as given in \refEq{eqn:adrc_discrete_ABLESO}.

For ADRC variants with transformed state variables, these updated matrices have to be turned into their transformed counterparts $\Matrix{\tilde{A}}_\mathrm{ESO}$, $\Matrix{\tilde{B}}_\mathrm{ESO}$ and $\Matrix{\tilde{L}}_\mathrm{ESO}$ via \refEq{eqn:adrc_opt_ABLESO}.

\subsubsection{Changing the plant model}

Changing $b_0$ leads to new values for $\Matrix{B}_\mathrm{d}$ (cf.\ \refEq{eqn:zoh_transform}) and $\Matrix{B}_\mathrm{ESO}$ (cf.\ \refEq{eqn:adrc_discrete_ABLESO}). It can be shown that updating $\Matrix{B}_\mathrm{d}$ and $\Matrix{B}_\mathrm{ESO}$ can be done by applying a scaling factor $b_0 / b_0^\mathrm{old}$:
\begin{equation}
\begin{aligned}
\Matrix{B}_\mathrm{d} &= \left( b_0 / b_0^\mathrm{old} \right) \cdot \Matrix{B}_\mathrm{d}^\mathrm{old}  \\
\Matrix{B}_\mathrm{ESO} &= \left( b_0 / b_0^\mathrm{old} \right) \cdot \Matrix{B}_\mathrm{ESO}^\mathrm{old}  \\
\end{aligned}
\label{eqn:change_parameter_plant_b}
\end{equation}

For the ADRC variants with transformed state variables, changing $b_0$ also results in a different  transformation matrix $\Matrix{T}^{-1}$ according to \refEq{eqn:adrc_opt_tinv}. In these cases, the observer matrices $\Matrix{\tilde{A}}_\mathrm{ESO}$, $\Matrix{\tilde{B}}_\mathrm{ESO}$ and $\Matrix{\tilde{L}}_\mathrm{ESO}$ have to be updated via \refEq{eqn:adrc_opt_ABLESO}. However, changing $b_0$ only results in a scaling factor ($\Matrix{T}^{-1} \cdot \Matrix{T}^\mathrm{old} = b_0^\mathrm{old} / b_0$). The influence of $b_0$ on $\Matrix{\tilde{A}}_\mathrm{ESO}$ and $\Matrix{\tilde{B}}_\mathrm{ESO}$ is cancelled out. Therefore it is sufficient to rescale $\Matrix{\tilde{L}}_\mathrm{ESO}$ and the state vector:
\begin{equation}
\begin{aligned}
\Matrix{\tilde{L}}_\mathrm{ESO} &= \left( b_0^\mathrm{old} / b_0 \right) \cdot \Matrix{\tilde{L}}_\mathrm{ESO}^\mathrm{old}  \\
\Vector{\tilde{x}}(k-1) &= \left( b_0^\mathrm{old} / b_0 \right) \cdot \Vector{\tilde{x}}^\mathrm{old}(k-1)  \\
\end{aligned}
\label{eqn:change_parameter_plant_lx}
\end{equation}

In order to keep the influence of the disturbance cancellation constant, the last element of the state vector (i.\,e.\ the second element in the first-order case and the third element in the second-order case) has to be scaled in the opposite manner:
\begin{equation}
\hat{x}_{2/3}(k-1) = \left( b_0 / b_0^\mathrm{old} \right) \cdot \hat{x}_{2/3}^\mathrm{old}(k-1)
\label{eqn:change_parameter_plant_xend}
\end{equation}

\refEq{eqn:change_parameter_plant_xend} also applies to the variants with transformed state variables, and therefore partially reverses the scaling performed with \refEq{eqn:change_parameter_plant_lx} in order to keep $\tilde{x}_{2/3}(k-1)$ constant.

Finally, $\Delta u(k-1)$ has to be adjusted for the incremental ADRC variants, again using \refEq{eqn:controller_init_deltau}.


\section{Examples}

The aim of this section is to demonstrate the various aspects of a modified practical ADRC implementation as introduced in sections \ref{sec:incremental}, \ref{sec:limitation} and \ref{sec:bumpless} in a visually intuitive manner. All experiments in this section are being carried out with a simulated plant and a discrete-time ADRC implementation.

As an example from the field of power electronics, a buck converter in peak current-mode control will be examined. ADRC will be employed for the outer voltage loop. A model for this plant, which will be used for the simulations in this section, can be found in \cite{Ridley:1991}. It is given as follows, with $v_\mathrm{o}$ being the output voltage of the converter and $i_\mathrm{c}$ the reference peak current value of the current loop:
\begin{align}
&\frac{v_\mathrm{o}(s)}{i_\mathrm{c}(s)} = K \cdot R \cdot  \frac{ \left(1 + s R_\mathrm{ESR} C\right) }{ \left(1 + s K R C\right) \cdot \left(1 + \D\frac{s}{\omega_\mathrm{n} Q} + \frac{s^2}{\omega_\mathrm{n}^2} \right) }
\\
& \text{with} \quad
K = \frac{1}{1 + \frac{R}{L \omega_\mathrm{n} Q}}
\quad \text{and} \quad
\omega_\mathrm{n} = \frac{\pi}{T_\mathrm{Switch}} = \frac{\pi}{T_\mathrm{Sample}} \notag
\end{align}

The parameters of the converter are $L = 1\,\U{mH}$, $C = 20\,\U{\mu F}$, $R_\mathrm{ESR} = 10\,\U{m\Omega}$, $R = 100\,\U{\Omega}$, with an input voltage of $400\,\U{V}$ and a nominal output voltage of $250\,\U{V}$. It is assumed that the slope compensation in the current loop is chosen such that $Q = 1$ holds, cf.\ \cite{Ridley:1991}.

Analysing the poles and zeros of the plant reveals that it is dominated by a first-order low-pass behaviour, which calls for the first-order case of ADRC. All that needs to be known for ADRC is $b_0$, which can be obtained according to \refSec{sec:adrc_continuous} from the DC gain and the dominant time constant as follows: $b_0 = \frac{KR}{KRC} = \frac{1}{C}$. This is a major simplification for the user compared to the modelling effort traditionally needed for the controller design for this plant \cite{Hallworth:2012}.

The discrete-time controller (with $T_\mathrm{sample} = 10\,\U{\mu s}$, corresponding to a switching frequency of the converter of $100\,\U{kHz}$) is parameterised according to \refSec{sec:adrc_parameter} for a closed-loop settling time $T_\mathrm{settle} = 2\,\U{ms}$, and the observer poles are placed with $k_\mathrm{ESO} = 5$. The control signal, i.\,e.\ the peak current reference value, will be limited to the interval $[0\,\U{A}, 5\,\U{A}]$. In order to make the effect of measurement noise on the control signal visible, normally distributed measurement noise (standard deviation $\sigma = 1\,\U{V}$) is added to the simulated converter output voltage.


\subsection{Rate Limitation}

\begin{figure}
    \centering%
    \includegraphics{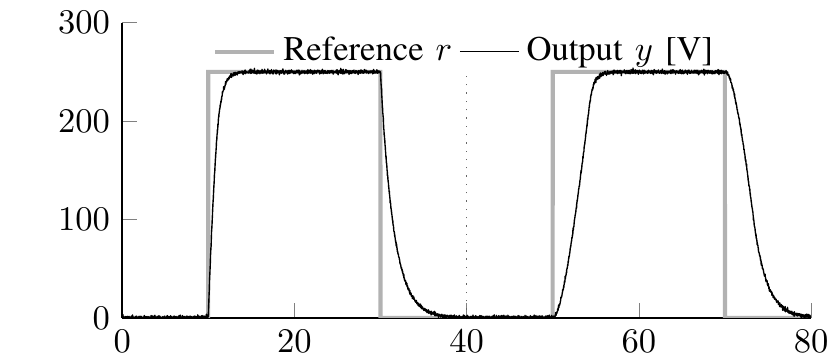}
    \includegraphics{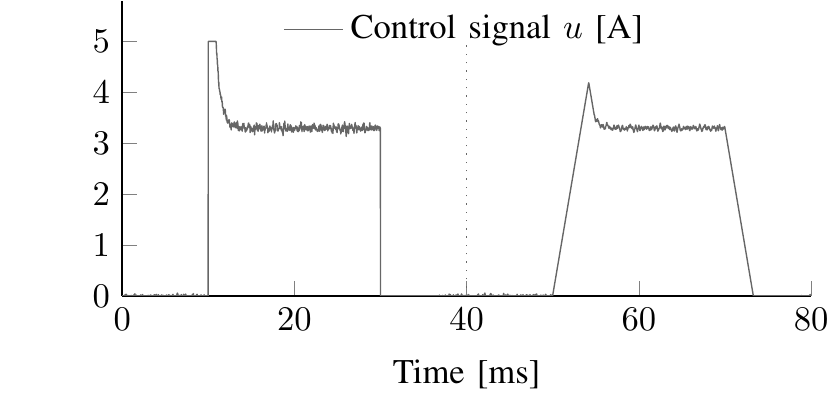}
    \caption{Example~A: Rate limitation of the control signal $u$ (enabled at $t = 40\,\U{ms}$). Controller, observer and limits are parameterised to make the effect of rate limitation ($\Delta u_\mathrm{max} = 1\,\U{A/ms}$) clearly visible. Note that transients down to a lower voltage are slower since the output capacitor of the converter cannot be actively discharged.}
    \label{fig:example_limit}
\end{figure}

In this first experiment, the effect of a rate limitation of the control signal will be demonstrated. To that end, a reference trajectory is being followed twice in the simulation results presented in \refFig{fig:example_limit}, once without and once with enabled rate limitation. Here, a rather strong rate limitation of $1\,\U{A/ms}$ (maximum allowed increase/decrease of the converter's peak current) was chosen to clearly demonstrate the effect on both the closed-loop dynamics as well as the control signal. Even the influence of measurement noise on the control signal is attenuated, without impacting the stability of the control loop.


\subsection{Bumpless Transfer to ADRC}

\begin{figure}
    \centering%
    \includegraphics{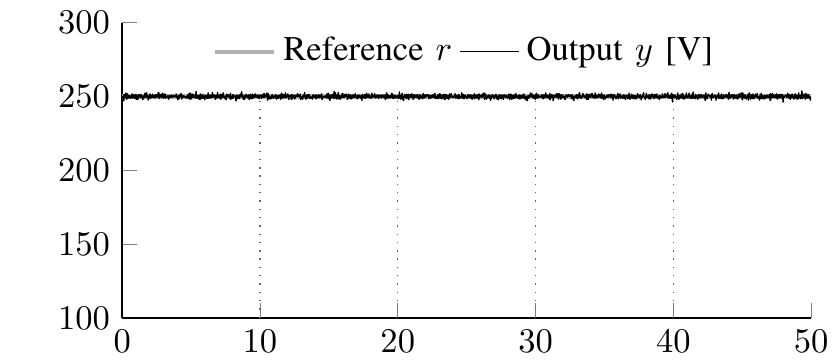}
    \includegraphics{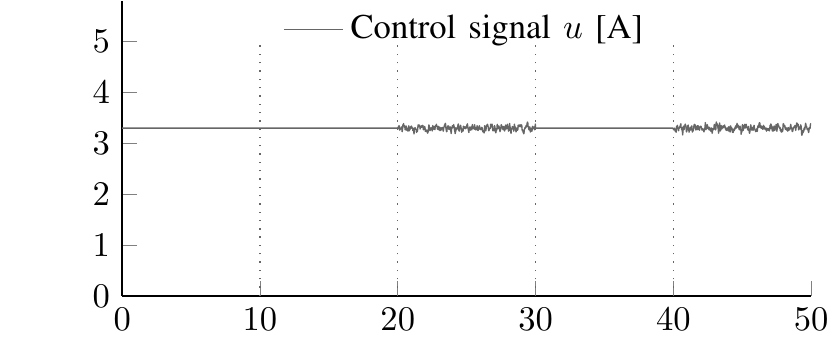}
    \includegraphics{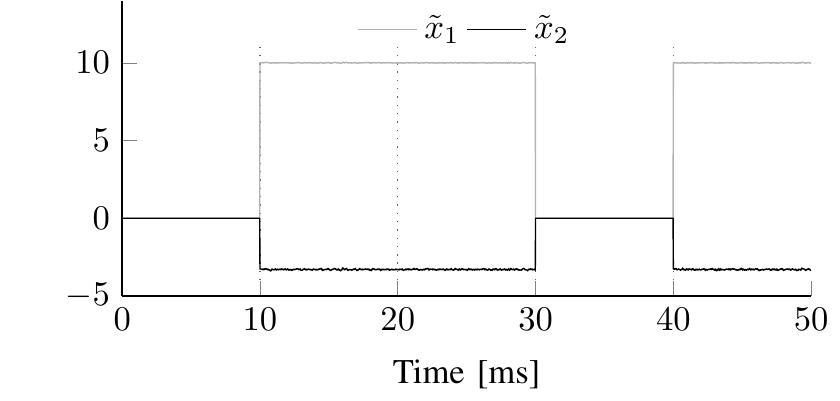}
    \caption{Example~B: Bumpless transfers from manual mode (in stationary state) to ADRC at $t = 20\,\U{ms}$ and $t = 40\,\U{ms}$. From $t = 30\,\U{ms}$ to $t = 40\,\U{ms}$, both controller and observer are disabled again. For the first transfer, the observer was enabled before at $t = 10\,\U{ms}$. For the second transfer at $t = 40\,\U{ms}$, both observer and controller are enabled jointly according to \refSec{sec:bumpless_enable_eso} and \refSec{sec:bumpless_enable_controller}. Note that practically no settling time is required for the state variables due to their initialisation as proposed in \refSec{sec:bumpless_enable_eso}.}
    \label{fig:example_bumpless_enable}
\end{figure}

Enabling a controller in a bumpless manner is another important aspect for a practical implementation. As shown in \refSec{sec:bumpless_enable_eso} and \refSec{sec:bumpless_enable_controller}, bumpless transfer from manual mode to ADRC is almost trivial and requires very few steps.

In the experiment presented in \refFig{fig:example_bumpless_enable}, the procedure of enabling ADRC was performed twice. At all times, the converter was in a stationary state at $250\,\U{V}$ output voltage. At first, the procedure was explicitely splitted in two phases (enabling the observer and then the actual control law). Then, after switching back to manual mode, both observer and controller were enabled at the same time. As expected, no bumps occur in the control signal or the process output in \refFig{fig:example_bumpless_enable} if $r = y$ holds when enabling the controller. Enabling the observer before the controller is recommended if the transfer does not take place in a stationary state.


\subsection{Bumpless Switch Between Incremental Control Laws}

\begin{figure}
    \centering%
    \includegraphics{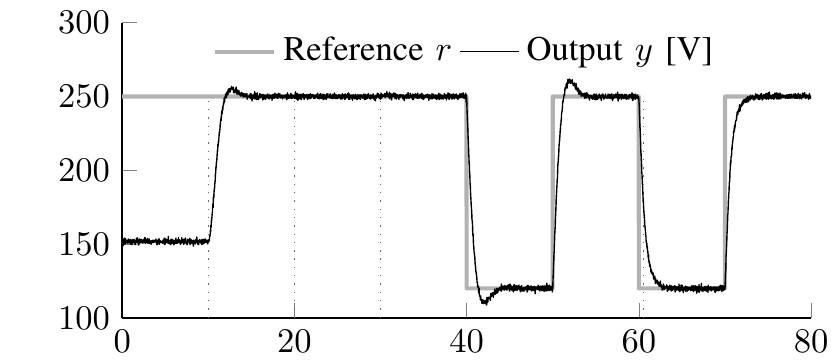}
    \includegraphics{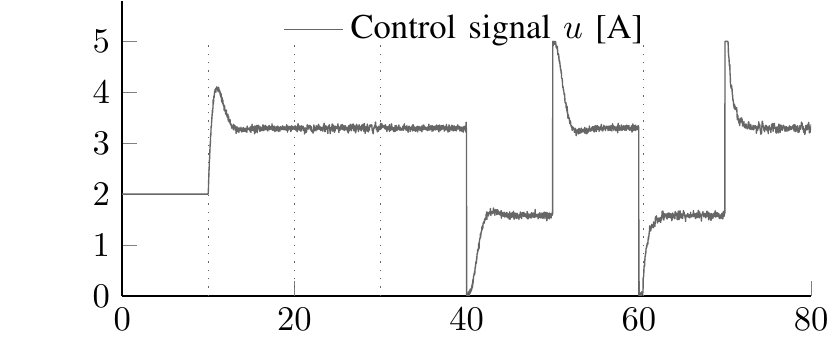}
    \includegraphics{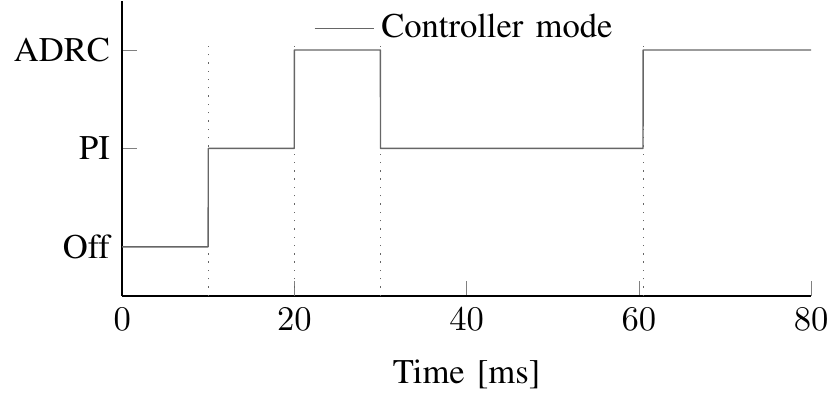}
    \caption{Example~C: Bumpless transfers between incremental PI control and incremental ADRC as described in \refSec{sec:incremental}. The transfers take place both in a stationary state (at $t = 20\,\U{ms}$ and $t = 30\,\U{ms}$) as well as during a transient (at $t = 60.5\,\U{ms}$). The PI control loop is deliberately parameterised for some overshoot to make the change in control loop behaviour more visible.
    }
    \label{fig:example_bumpless_pi}
\end{figure}

An important feature of an incremental controller implementation is that one can easily switch between different control laws while avoiding jumps in the control signal. Similar to transferring from manual mode to ADRC, one only has to follow the few steps described in \refSec{sec:bumpless_enable_eso} and \refSec{sec:bumpless_enable_controller} when switching from a different control law to incremental ADRC.

In this experiment, switching from manual mode to incremental PI control followed by switching to incremental ADRC (and back) will be demonstrated, both in a stationary state and during a transient. Both controllers operate using a common limiting integrator as depicted in \refFig{fig:limiter_inc}. The results are presented in \refFig{fig:example_bumpless_pi} and confirm the expected smooth transition under these conditions.


\subsection{Bumpless Parameter Changes}

\begin{figure}
    \centering%
    \includegraphics{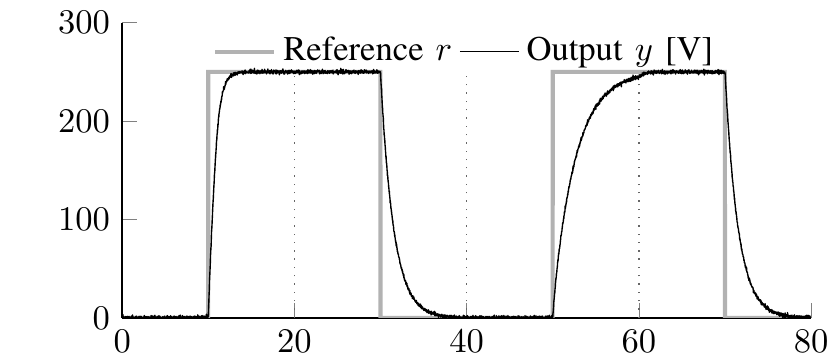}
    \includegraphics{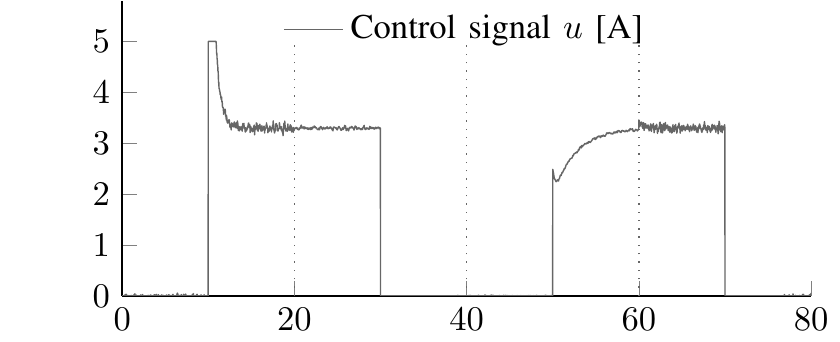}
    \includegraphics{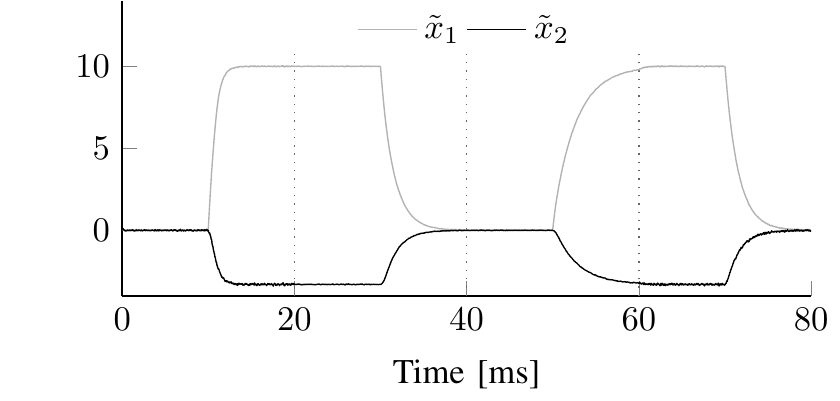}
    \caption{Example~D: Performing bumpless parameter changes according to \refSec{sec:bumpless_parameter}. At $t = 20\,\U{ms}$ the observer bandwidth is reduced by 50\,\%. From $t = 40\,\U{ms}$, the controller is adjusted for a 400\,\% increased closed-loop settling time. At $t = 60\,\U{ms}$, all parameters are set back to their initial values.}
    \label{fig:example_bumpless_parameter}
\end{figure}

As a final experiment, bumpless changes in the controller and observer parameters according to the guidelines developed in \refSec{sec:bumpless_parameter} will be demonstrated. While the control loop is  required to follow a series of reference signal steps as shown in \refFig{fig:example_bumpless_parameter}, the observer bandwidth is reduced by 50\,\%, resulting in a less noisy control signal and state estimates as well as a slightly degraded controller performance. Subsequently, the closed-loop poles are moved for a slower response, and finally both controller and observer parameters are reset to their initial values simultaneously. As expected, no jumps are visible in the state estimates and in the control signal if these parameter changes are performed when $r = y$ holds.


\section{Conclusion}

This article covered different aspects of discrete-time ADRC that can be very important in a practical implementation, but are often overlooked in theoretical studies. Firstly, a novel incremental form of discrete-time ADRC was introduced, which can be seamlessly integrated with other incremental control algorithms or be used for actuators with integral behaviour.
Secondly, means of limiting the ADRC control signal magnitude and/or rate are given. Especially rate limitation is feature rarely found in common controllers and novel to ADRC, but can provide benefits such as guaranteed rise times for the control signal after reference or disturbance steps.
A third part revolves around the notion of bumpless changes to the controller, i.\,e.\ enabling the controller or performing parameter changes while preventing jumps in the control signal.

With this contribution, the ADRC concept has been strengthened with features that can be essential in an industrial setting. Practitioners are now enabled to implement non-incremental or incremental variants of discrete-time ADRC with or without limitations in the control signal, and perform bumpless changes to the controller and its parameters.


\end{document}